\documentstyle[12pt]{article}

\newcommand{\be}{\begin{equation}}
\newcommand{\ee}{\end{equation}}
\newcommand{\teta}{\theta}
\newcommand{\p}{\partial}
\newcommand{\beqn}{\begin{eqnarray}}
\newcommand{\eeqn}{\end{eqnarray}}
\newcommand{\nn}{\nonumber}

\newcommand{\f}{\frac}

\newcommand{\r}{\right}
\newcommand{\lf}{\left}
\newcommand{\bg}{\bigg}
\newcommand{\bfl}{\begin{flushleft}}
\newcommand{\efl}{\end{flushleft}}
\newcommand{\br}{\begin{flushright}}
\newcommand{\er}{\begin{flushright}}

\begin{document}
\begin{titlepage}
\flushright{IP/BBSR/2002-26}
\flushright{MRI-P-021101}
\flushright{hep-th/0211115}

\vspace{1in}

\begin{center}
\Large
{\bf Coincident Dp-branes in codimension two}

\vspace{1in}

\normalsize

\large{ Shesansu Sekhar  Pal\footnote
{\rm  shesansu@iopb.res.in}}\\


{\em  Institute of Physics \\
Bhubaneswar - 751005, India }

\large{Sudhakar Panda\footnote{\rm panda@mri.ernet.in}\\
{\em Harish-Chandra Research Institute,\\
 Allahabad, 211019, India \\}}

\end{center}

\vspace{1in}

\baselineskip=24pt
\begin{abstract}
We derive the mass shell condition 
for N coincident D0 branes in codimension two 
i.e. in space-time dimension three. Using this we present the action for 
this system in first order formalism. Our analysis is 
restricted to flat space-time.

\end{abstract}

\end{titlepage}
\section{Introduction}
In recent years the importance  of Dp branes have been realized through the 
studies of black hole entropy, AdS/CFT correspondence, tachyon 
condensation etc. Hence, it has 
been necessary to study the dynamics of these branes in various configurations. 
The dynamics of a single brane (both BPS and non-BPS) is known to be 
described by the sum of a Born-Infeld type action and a Chern-Simon action. 
The various degrees of freedom described by such an action include
a $U(1)$ valued  gauge field, transverse scalars, induced metric, induced 
Kalb-Ramond field on the world volume of the Dp brane coming from the NS-NS 
sector, and  antisymmetric field of various ranks coming from the R-R 
sector. 
The supersymmetric and kappa symmetric invariant action describing a 
single BPS D-brane in flat space is known for quite some time \cite{aps} and 
in curved space, in particular in Type II background, \cite{cgnwsbt}. 
The simplest 
generalization of this is the system of N number of coincident Dp branes, 
instead of a single brane. It is known that for such a system one gets a 
non-abelian theory with a gauge group $U(N)$\cite{ew} and the 
fields that live on the brane take values in the adjoint representation of 
$U(N)$. It is also known that these branes are charged under R-R 
fields \cite{jp} and one expects that this system should also preserve 
half of the supersymmetries just like a single BPS brane. However, to show 
that this is indeed true one needs to show that the world volume action is 
in fact invariant under kappa symmetry as in the case for a single BPS 
Dp-brane. But such a proof has not been achieved so far. The main obstacle 
is that unlike a single brane where the kappa symmetry is realized 
as an abelian symmetry, for the case of multiple coincident branes, this 
symmetry is enhanced to a non-abelian symmetry.  \\
 
The dynamics of the N coincident Dp branes is described in great detail 
in \cite{rm}, which is derived by starting with a space filling brane but 
subsequent T-dualisation gives the desired action.
The bosonic parts of the world volume action for N coincident Dp-branes are 
given as: 
\be
\label{bi_action}  
S_{BI}=-T_p \int d^{p+1}\sigma STr\left( e^{-\phi}{\sqrt{ -det(P[E_{ab}
+E_{ai}(Q^{-1}-\delta)^{ij}E_{jb}]+\lambda F_{ab})det(Q^i_j)}}\right),
\ee 
with
\be     
E_{\mu\nu}=G_{\mu\nu}+B_{\mu\nu},~~~~~~and~~~~~ 
Q^i_j=\delta^i_j+i\lambda [\Phi^i,\Phi^k]E_{kj},
\ee
and 
\be
\label{cs_action}
S_{CS}=\mu_p\int STr\lf( P[e^{i\lambda i_{\Phi}i_{\phi}}(\sum_n C^{(n)}e^B)]
e^{\lambda F}\r),
\ee

where $G_{\mu\nu}$, $B_{\mu\nu}$, $\phi$, are the metric, Kalb-Ramond and the 
dilaton field coming from the NS-NS sector, respectively. $F_{ab}$ is the 
$U(N)$ valued field strength that live on the world volume of the Dp brane and 
$C^{(n)}$'s are the rank $n$ antisymmetric field coming from R-R sector. 
$P$ is the pull-back which acts on the  bulk fields and brings it onto the 
world 
volume of the brane. $\Phi^i$'s are  the $U(N)$ valued scalars, and STr 
stands for the symmetrised average over all orderings of the $U(N)$ valued 
objects and then taking the trace.  The pull back is defined as 
\be
\label{pull_back}
P[E]_{ab}=E_{ab}+E_{ ai}D_{b}\Phi^i+E_{ib}D_a\Phi^i+D_a\Phi^iD_b\Phi^j E_{ij}
\ee
where $D_a$ are the covariant derivatives with respect to the world volume 
coordinates and the gauge fields being the connection. The expression of 
the pull-back, as written in 
eq.(\ref{pull_back}), is expressed in the 
 static gauge choice, namely  $X^a=\sigma^a$, and 
$X^i=\Phi^i$, with the choice $\lambda=1$, i.e. $p+1$ coordinates of the 
target space are identified with the world volume coordinates $\sigma^a$ and 
the indices i, j denotes the target space coordinates 
transverse to the Dp branes.
Hence the space-time symmetry is now broken to world volume symmetry times 
the symmetry in the transverse directions to the branes. Since the above 
action is written in the static gauge, it implies that it cannot be world 
volume reparametrisation invariant and invariant under the full target 
space 
diffeomorphism, except for the case of space filling branes. In order to 
write an action which is both supersymmetric and $\kappa$ symmetric invariant,
we have to  make it both world volume reparametrisation invariant and target 
space diffeomorphism invariant, otherwise it might be difficult to write 
down an 
action which is both supersymmetric and $\kappa$ symmetric invariant. 
In a recent paper Sorokin has proposed a first order action
for coincident D0 branes in codimension one which is shown to have both 
the world volume and space-time diffeomorphism invariance \cite{ds}. In 
this letter we extend this proposal to coincident D0 branes for 
codimension two and write down the bosonic part of the action.\\

Before we review the above construction, it is useful to note an 
interesting point about the Chern-Simon action, namely the N
coincident Dp branes  can 
couple to higher rank R-R fields through $e^{ii_{\Phi}i_{\Phi}}$, \cite{rm}, 
and which in turn gives rise to various fuzzy surfaces, like $S^2$ and $S^4$ 
\cite{nro} and also studied in \cite{dtv}.\\

In \cite{bds} a possible supersymmetric and kappa symmetric invariant action 
has been constructed but up to order $F^2$  in the field strength and in 
\cite{wv} the supersymmetric and fermion couplings have been considered in 
the context of Matrix theory of D0 branes.

The construction of \cite{ds} is to consider coincident N Dp branes as a 
single brane configuration, and call it NDp-brane. The transverse coordinates 
of this single brane is described by $x^i(\sigma)$, which is given as the 
trace of the $U(N)$ valued scalars $\Phi^i$, i.e.
\be
x^i(\sigma)=\f{1}{N}Tr\Phi^i(\sigma).
\ee
Now this transverse coordinate $x^i$ together with the world volume coordinate
$\sigma^a$ represents the coordinates of the single brane in the target 
space-time, i.e. $x^{\mu}(\sigma)=(\sigma^a, x^i(\sigma))$ with $\mu=0,1,
\ldots,D-1$, in a
D dimensional target space in the static gauge.  In order to make the world 
volume theory of that single brane diffeomorphism invariant, let's introduce 
p+1 coordinates, $x^a(\sigma)$,  as the world volume coordinates of the 
(single) NDp-brane, i.e. 
\be
\label{decomposition}
x^{\mu}(\sigma)=(x^a(\sigma), x^i(\sigma)).
\ee 
Then the $U(N)$ vector fields $A_a(\sigma)$ and the traceless scalars 
\be
\label{u_n=u_1+su_n}
\phi^i(\sigma)\equiv\Phi^i(\sigma)-x^i(\sigma)I, ~~~~\in~~SU(N),
\ee
takes values in $SU(N)$, and are being considered as pure world volume vector 
fields and scalar fields living on the NDp-brane. At this stage, we are 
assuming that this  single NDp-brane
action is invariant under only one kappa symmetry as rest N-1 kappa 
symmetries  have been gauge fixed by construction. This feature will be 
more explicit in the next section. \\

The action of N coincident Dp-branes, eq.(\ref{bi_action}), gives us 
a complicated form after imposing the world volume reparametrisation 
invariance in a generic background. Hence, to avoid complications, we shall 
deal with the action in 
co-dimension one and two in flat spacetime, as the mass 
shell condition becomes complicated for higher codimension. The meaning of 
co-dimension one and two is that the dimension of the target space for a 
single NDp-brane becomes $p+2$, and $p+3$ respectively.\\

In section 2 we shall write down the world volume reparametrisation invariant 
action of N Dp branes in co-dimension $d$ in flat spacetime, and   
in section 3 we shall write down the world volume reparametrisation 
invariant and kappa 
symmetric invariant action of N D0 branes in co-dimension 1 by implementing 
the proposal of \cite{ds}. In section 4 we shall write down the bosonic part
of the action in 
co-dimension 2 i.e. in space-time dimension 3. Then we shall conclude in 
section 5 and in the appendix we have derived the mass shell condition for 
the co-dimension two case.
\section{N Dp branes in codimension d}
In codimension d the Born-Infeld action  eq. (\ref{bi_action})
 becomes, in the flat background, i.e. $B_{\mu\nu}=0$, $G_{\mu\nu}=
\eta_{\mu\nu}$, $\phi=0$, and  in the static gauge
\beqn
\label{bi+cs_action}
S=-T_p\int d^{p+1}\sigma STr& &\bg[(-det(\p_a\sigma^c \p_b\sigma^d\eta_{cd}+
D_a\Phi^iD_b\Phi^j\eta_{ij}\nn \\& &+D_a\Phi^k D_b\Phi^l\eta_{ki}(Q^{-1}-
\delta)^{ij}\eta_{lj}+F_{ab}))det(Q^i_j)\bg]^{\f{1}{2}},
\eeqn
with 
\be
Q^i_j=\delta^i_j+i[\Phi^i,\Phi^k]\eta_{kj}.
\ee
Let's restore the world volume reparametrisation invariance by writing 
$\sigma^a=x^a(\sigma)$, using eq. (\ref{decomposition}) and eq. 
(\ref{u_n=u_1+su_n}) and substituting it in eq.(\ref{bi+cs_action}). This 
gives us the following action:
\beqn
\label{bi+cs_rep_invariant_action}
S=-T_p\int d^{p+1}\sigma STr& &\bg[(-det(\p_a x^{\mu} \p_b x^{\nu}
\eta_{\mu\nu}+\p_ax^{\mu}D_b\phi^j\eta_{\mu j}+D_a \phi^i
\p_b x^{\nu}\eta_{i\nu}\nn \\& &+D_a\phi^iD_b\phi^j\eta_{ij}+D_a\phi^k D_b
\phi^l L_{kl}+\p_a x^{\mu}\p_b x^{\nu} L_{\mu\nu}\nn \\& &+\p_a x^{\mu}D_b
\phi^l L_{\mu l}+D_a \phi^k\p_b x^{\nu} L_{k\nu}
+F_{ab}))det(Q^i_j)\bg]^{\f{1}{2}},
\eeqn
where $L_{AB}=\eta_{Ai}(Q^{-1}-\delta)^{ij}\eta_{jB}$, and A, B can take 
values (a,i). The above action has a complicated form and is not suitable 
for further analysis i.e. to write down a supersymmetric and $\kappa$
symmetric invariant action of N Dp-branes in co-dimension $d$. So, we 
consider the simpler cases of co-dimension one and two only.
\section{N D0 branes in codimension one}
The bosonic part of the action that follows from eq. 
(\ref{bi+cs_rep_invariant_action})
for N D0 branes in codimension one, i.e. in space-time dimension two, is
\be
\label{nd0_action}
S=-T_0\int d\tau STr\sqrt{-[{\dot x^{\mu}}{\dot x^{\nu}}\eta_{\mu\nu}+
2{\dot x^1}{\dot\phi^1}+({\dot\phi^1})^2]},
\ee
where $T_0$ is the tension of a D0 brane, $\tau$ is the world line 
parameter and $\cdot $ denotes the derivative with respect this parameter. 
It is very easy to see 
that the action in eq. (\ref{nd0_action}) is world line reparametrisation 
invariant, hence, the corresponding first 
class constraint would be some generalization of the mass shell condition for 
$\phi=0$. \\

Let $p_{\mu}$ be the momentum associated to $x^{\mu}$. Before, we start to 
write down the canonical momenta associated to $x^{\mu}$ and $\phi$, we 
rewrite the action eq.( \ref{nd0_action}) in a suggestive manner.
\be
S=-T_0\int d\tau STr \sqrt{({\dot x^{0}})^2-({\dot\Phi})^2},
\ee
where $\Phi=x^1 I+\phi^1$, from eq. (\ref{u_n=u_1+su_n}), and the momenta 
associated to $x^0$ and $\Phi$ are
\be 
p_0=-T_0{\dot x^0}STr\bg[\sqrt{({\dot x^{0}})^2-({\dot\Phi})^2}\bg]^{-1},
\ee
and
\be
p_{\Phi}=T_0{\dot\Phi}\bg[\sqrt{({\dot x^{0}})^2-({\dot\Phi})^2}\bg]^{-1}.
\ee
The momentum $p_{\Phi}=\f{1}{N}p_1 I+p_{\phi}$, with $p_{\Phi}\in U(N)$ and 
$p_{\phi}\in SU(N)$, where
\be
p_1=T_0STr~[{\dot\Phi}(\sqrt{({\dot x^{0}})^2-({\dot\Phi})^2})^{-1}].
\ee 
The mass shell condition for N D0 brane in  spacetime dimension two is
\be
\label{mass_shell_condition_1}
-(p_0)^2+\lf(STr[\sqrt{p^2_{\Phi}+T^2_0 I}]\r)^2=0,
\ee 
and this reduces to the usual mass shell condition for N D0 branes when 
$\phi=0$. Hence, the first order bosonic part of the action is
\be
\label{nd_bosonic}
S=STr\int d\tau\bg[\f{1}{N}p_{\mu}{\dot x^{\mu}}+p_{\phi}{\dot\phi}-
\f{e(\tau)}{2N}[-(p_0)^2+\lf(STr[\sqrt{p^2_{\Phi}+T^2_0 I}]\r)^2]\bg],
\ee  
where $e(\tau)$ is the Lagrange multiplier for the mass shell condition as 
given in eq. (\ref{mass_shell_condition_1}).\\

We can now supersymmetrise the above bosonic action by introducing a two 
component Majorana spinor field 

\beqn
\teta^{\alpha}= \left(\begin{array}{cc} \teta^1 \\ \teta^2 \end{array}\right),
\eeqn
for $\phi=0$. In the first order formalism, we know the form of  the 
supersymmetric invariant action of a super particle with mass m 
and it is given by
\be
\label{bosonic_susy_action}
S=\int d\tau ~~[p_{\mu}({\dot x^{\mu}}+i {\bar\teta }\gamma^{\mu}{\dot\teta})-
\f{e(\tau)}{2}(p_{\mu}p^{\mu}+m^2)+m{\bar\teta}\gamma^2{\dot\teta}],
\ee
where ${\bar\teta}=\teta^T\gamma^0$ and the form of the $\gamma$ matrices are
\beqn
\gamma^0  = \left( 
\begin{array}{cc} 0 & -i \\
                   i & 0  \end{array} 
\right),~~~\gamma^1= \left( \begin{array}{cc} 0 & i \\ i & 0 \end{array} 
\right),~~~\gamma^2=\gamma^0\gamma^1=\left( \begin{array}{cc} 1 & 0 \\ 0 & -1 
\end{array} \right).
\eeqn
The corresponding Clifford algebra is
\be
\{\gamma^{a},\gamma^{b}\}=-2\eta^{ab} ~~~{\rm with}~~~ \eta^{ab}=(-,+). 
\ee

The action eq. (\ref{bosonic_susy_action}) is invariant under both global 
supersymmetry and local $\kappa$-symmetry transformations, and the form of 
transformations are
\be
\label{susy_transf}
\delta_{\epsilon}\teta=\epsilon, ~~~\delta_{\epsilon}x^{\mu}=-i{\bar\epsilon}
\gamma^{\mu}\teta, ~~~\delta_{\epsilon}(others)=0,
\ee
\be
\delta_{\kappa}x^{\mu}=i\delta_{\kappa}{\bar\teta}\gamma^{\mu}\teta,~~~
\delta_{\kappa}e=4i{\bar\kappa}{\dot\teta},~~~\delta_{\kappa}\teta=
(\gamma^{\mu}p_{\mu}-im\gamma^2)\kappa,~~~\delta_{\kappa}(others)=0.
\ee
Lets supersymmetrise the action eq. (\ref{nd_bosonic}) for the N coincident 
Dp-branes in codimension one for $\phi\neq 0$ and introducing the fermions 
$ \psi$, following the action of a super particle as written in 
eq. (\ref{bosonic_susy_action}), we get  

\beqn
\label{action_cod_1}
S&=&STr\int d\tau\bigg(\f{1}{N}p_{\mu}({\dot x^{\mu}}+i{\bar\teta}\gamma^{\mu}
{\dot\teta})+p_{\phi}{\dot\phi}-\f{e}{2 N}[p_{\mu}p^{\mu}+\f{M^2(p_{\phi},p_1)}{N}]+\f{M(p_{\phi},p_1)}{N}{\bar\teta}\gamma^2{\dot\teta}\nn \\
& &-i{\bar\psi}{\dot\psi}\bigg),
\eeqn
where $\psi~~\in~~SU(N)$ and $M^2(p_{\phi}, 
p_1)= (STr\sqrt{p^2_{\Phi}+T^2_0 I})^2-p^2_1$. Where the supersymmetry
and the kappa symmetry transformations are given by
\be 
\delta_{\epsilon}\teta=\epsilon, ~~~\delta_{\epsilon}x^{\mu}=-i{\bar\epsilon}
\gamma^{\mu}\teta-{\bar\epsilon}\gamma^2\teta \f{\p M}{\p p_1}\delta^{\mu}_1,~~~\delta_{\epsilon}\phi=-{\bar\epsilon}\gamma^2\teta\f{\p M}{N\p p_{\phi}},
 ~~~\delta_{\epsilon}(others)=0,
\ee
\beqn 
& &\delta_{\kappa}x^{\mu}=i\delta_{\kappa}{\bar\teta}\gamma^{\mu}\teta+\delta_{\kappa}{\bar\teta}\gamma^2\teta \f{\p M}{\p p_1}\delta^{\mu}_1,~~~
\delta_{\kappa}e=4i{\bar\kappa}{\dot\teta},~~~\delta_{\kappa}\teta=
(\gamma^{\mu}p_{\mu}-i\f{M(p_{\phi}, p_1)}{N}\gamma^2)\kappa,\nn \\ 
& & ~~~\delta_{\kappa}\phi=\delta_{\kappa}{\bar\teta}\gamma^2\teta \f{\p M}{N \p p_{\phi}}, ~~~
\delta_{\kappa}(others)=0, 
\eeqn
\section{N DO branes in codimension two}
The bosonic part of the action that governs the dynamics of N D0 branes 
in codimension two is given by
\be
S =-T_0 STr\int d\tau~\bg[\sqrt{({\dot 
x^0})^2(detQ_{ij})-(detQ_{ij}){\dot\Phi^k}{\dot\Phi^l}Q^{-1}_{kl}}\bg], 
\ee
where i, j of  determinant  $Q_{ij}$ can take  only two values, namely, 1, 2. 
Evaluating the determinant in three space-time dimensions, give rise to
\be 
\label{action_for_codimension_2}
S =-T_0 STr\int d\tau \sqrt{\lf({\dot {x^0}^2}-{\dot {x^1}^2}-{\dot 
{x^2}^2} -{\dot{\phi^1}^2}-{\dot{\phi^2}^2}-2{\dot 
{x^1}}{\dot\phi^1}-2{\dot x^2}{\dot\phi^2}-{\dot {x^0}^2}[\phi^1,\phi^2]^2\r)},
\ee
using eq. (\ref{u_n=u_1+su_n}), we can rewrite the above equation as
\be 
S= -T_0 STr\int d\tau\sqrt{({\dot 
{x^0}^2}-{\dot{\Phi^1}^2}-{\dot{\Phi^2}^2}-{\dot {x^0}^2}[\Phi^1,\Phi^2]^2)}.
\ee
The momenta conjugate to $x^0, \Phi^1, \Phi^2$ are
\beqn
& &p_0=-T_0 {\dot x^0} STr \lf( \f{1-[\Phi^1,\Phi^2]^2}{\sqrt{{\dot 
{x^0}^2}- {\dot{\Phi^1}^2}-{\dot{\Phi^2}^2}- {\dot 
{x^0}^2}[\Phi^1,\Phi^2]^2}}\r),\nn \\& &
 p_{\Phi^1}=\f{T_0}{2}\Bigg\{{\dot\Phi^1}, \f{1}{\sqrt{{\dot {x^0}^2}- 
{\dot{\Phi^1}^2}-{\dot{\Phi^2}^2}- {\dot 
{x^0}^2}[\Phi^1,\Phi^2]^2}}\Bigg\},\nn \\& &
p_{\Phi^2}=\f{T_0}{2}\Bigg\{{\dot\Phi^2}, \f{1}{\sqrt{{\dot {x^0}^2}- 
{\dot{\Phi^1}^2}-{\dot{\Phi^2}^2}-{\dot {x^0}^2}[\Phi^1,\Phi^2]^2}}\Bigg\}.
\eeqn
The momenta $p_{\Phi^i}=\frac{1}{N}p_i I+p_{\phi^i}$, with
$p_{\Phi^i}\in$ U(N) and $p_{\phi^i}\in$ SU(N).
The mass shell condition for the N D0 brane in codimension two is (see the 
appendix for details)
\be
\label{mass_shell_condition_2}
-p^2_0+\bg[STr\sqrt{(p^2_{\Phi^1}+p^2_{\Phi^2}+T^2_0) 
(1-[\Phi^1,\Phi^2]^2)}\bg]^2=0.
\ee
It is easy to see that this mass shell condition reduces to the mass 
shell condition for codimension one, eq. (\ref{mass_shell_condition_1}), 
for either $\Phi^1=0, ~~{\rm or}~~ \Phi^2=0$. 

Let's rewrite eq. (\ref{action_for_codimension_2}) in the first order 
formalism  as 
\be
S =STr\int d\tau\bg[\f{1}{N}p_{\mu}{\dot x^{\mu}}+ 
p_{\phi^1}{\dot\phi^1}+p_{\phi^2}{\dot\phi^2}-\f{e(\tau)} 
{2N}(p_{\mu}p^{\mu}+{\cal M}^2(p_{\phi^1}, p_{\phi^2}, p_1, p_2, \phi_1, \phi_2) )\bg],
\ee
where ${\cal M}^2(p_{\phi^1}, p_{\phi^2}, p_1, p_2, \phi_1, \phi_2) 
=\lf(STr\sqrt{(p^2_{\Phi^1}+p^2_{\Phi^2}+T^2_0 I) (1-[\Phi^1, 
\Phi^2]^2)}\r)^2-p^2_1-p^2_2$.\\

One can follow the procedure in the previous section to obtain a 
supersymmetric and kappa symmetric invariant action, which we donot 
pursue here. On the other hand we note that the above action is not 
manifestly Lorentz invariant because of the 
presence of the last two terms in the expression of the square of the 
effective mass, eventhough, this action has been derived from Myers action 
which is perfectly Lorentz invariant. This isuue remains to be a puzzle 
for us and needs further investigation. 

\section{Conclusion}
In this note, we have examined Sorokin's  prescription to see 
the coincident N D0 branes as a single ND0-brane in different 
co-dimensions to get atleast the bosonic part of the  
action in the first order formalism. 
In particular, we have considered only  the N D0 branes  in 
co-dimension one and two  in flat spacetime. To implement the same 
prescription for higher branes and for higher codimensions one needs to 
know the corresponding mass shell conditions which become more 
complicated and the analysis becomes more involved (as seen in the 
appendix) though in principle it 
is possible. Moreover, even in this prescription we still encounter the 
usual problem of the nature of 'trace' one should adopt. However, the 
prescription by itself is beautiful since the kappa symmetry is reduced to 
an abelian symmetry, as we saw while in trying to write down the action of 
N coincident D0-branes in the first order formalism for codimension one case.
The main draw back of this prescription seems to be the abscence of 
manifest Lorentz 
invariance, but we can get back the invariance if we drop those terms from the 
square of the mass.
In writing down the action  we have to get pass 
two hurdles, first, the  correct mass shell condition and second the 
transformation rules 
of the fields keeping in mind that  ${\cal M}^2$ becomes a function of 
the nonabelian scalars, 
in order to make the action both supersymmetric and kappa symmetric invariant\footnote{We are hoping to come back to it in future for both supersymmetric and kappa symmetric invariant action.}. \\

\begin{flushleft}
{\bf Acknowledgement}
\end{flushleft}
We wish to thank Dmitri Sorokin for usuful email correspondences. 
SSP would like gratefully acknowledges the warm hospitality at the 
Harish-Chandra Research Institute, Allahabad, where part of the work 
is performed.
\section{Appendix A}
In this section we shall derive  the mass shell condition eq. (\ref
{mass_shell_condition_2}) by expanding it term by term. For this purpose 
we define some notations.
\beqn
& &1-[\Phi^1, \Phi^2]^2\equiv Z\equiv 1-B\nn, ~~\Rightarrow B = [\Phi^1, 
\Phi^2]^2,\nn  \\ & &
{\dot {\Phi^1}^2}+{\dot{\Phi^2}^2}\equiv Y,\nn \\& &
p^2_{\Phi^1}+p^2_{\Phi^2}\equiv A.
\eeqn
The momentum $p_0$ in these notations becomes
\be
p_0=-T_0 {\dot x^0} STr ~\lf(\f{Z}{\sqrt{{\dot {x^0}}^2 Z-Y}}\r),
\ee
and expanding it, we get
\be
\label{exp_of_p_0}
p_0=-T_0 STr~[Z^{\f{1}{2}}+\f{1}{2} \f{Z^{\f{-1}{2}}Y}{{\dot {x^0}}^2}+\f{3}{8}
\f{Z^{\f{-3}{2}}Y^2}{{\dot {x^0}}^4}+\ldots].\footnote{We have used the 
property
of symmetrised trace. Note: We shall also use the property of STr while 
expanding $ p^2_{\Phi^1}$ and $p^2_{\Phi^2}$.}
\ee
Now the other term present in the mass shell condition becomes, in these 
notations
\be
STr\sqrt{(p^2_{\Phi^1}+p^2_{\Phi^2}+T^2_0) Z}= STr\sqrt{(A+T^2_0)Z},
\ee 
which after expanding takes the form
\be
\label{exp_of_restterm}
T_0 STr Z^{\f{1}{2}}+\f{1}{2T_0} STr(Z^{\f{1}{2}}A)-\f{1}{8T^3_0} 
STr(Z^{\f{1}{2}}A^2)+\ldots.
\ee
The second term STr($Z^{\f{1}{2}}A)$ and the 3rd term STr($Z^{\f{1} 
{2}}A^2$) of the above equation can be written as
\beqn
& &\f{1}{2T_0}STr[ 
Z^{\f{1}{2}}A]=\f{T_0}{2{\dot{x^0}^2}}STr(Z^{\f{-1}{2}}Y) + 
\f{T_0}{2{\dot{x^0}^4}}STr(Z^{\f{-3}{2}}Y^2)+\ldots,\nn \\
& &-\f{1}{8T^3_0}STr(Z^{\f{1}{2}}A^2) 
=-\f{T_0}{8{\dot{x^0}^4}}STr(Z^{\f{-3}{2}}Y^2)+\ldots,
\eeqn
substituting these expressions of the second term and the third term in 
eq.(\ref{exp_of_restterm}), we see that it becomes exactly the -ve of 
the eq.(\ref{exp_of_p_0}). Which upon substituting in eq.
(\ref{mass_shell_condition_2}) gives us the right hand side.  Hence, 
the mass shell condition is proved.



\vspace{.7in}
\begin{center}
{\bf References}
\end{center} 
\begin{enumerate}

\bibitem{aps} M. Aganagic, C. Popescu and J.H. Schwarz, ``D-brane actions 
with local kappa symmetry,'' Phys. Lett. {\bf B 393} (1997) 311, 
[hep-th/9610249]; ``Gauge-invariant and gauge fixed D-brane actions,
'' Nucl. Phys. {\bf B 495} (1997) 99, [hep-th/9612080]; A.A. Tseytlin, 
``Born-Infeld action, supersymmetry and string theory,'' hep-th/9908105 and 
references therein.
\bibitem{cgnwsbt}  M. Cederwall, A. von Gussich, B. E.W. Nilsson and
 A. Westerberg, ``The Dirichlet Super-Three-Brane in Ten-Dimensional Type IIB Supergravity,'' Nucl. Phys. {\bf B490} (1997) 163, [hep-th/9610148];  M. Cederwall, A. von Gussich, B. E.W. Nilsson, P. Sundell and A. Westerberg, ``The Dirichlet Super-p-Branes in Ten-Dimensional Type IIA and IIB Supergravity,'' Nucl. Phys. {\bf B490} (1997) 179, [hep-th/9611159]; 
E. Bergshoeff and P.K. Townsend, ``Super D-branes,'' Nucl. Phys. {\bf B490} (1997) 145, [hep-th/9611173].
\bibitem{ew} E. Witten, ``Bound states of strings and p-branes,'' 
Nucl. Phys. {\bf B 460} (1996) 335, [hep-th/9510135].
\bibitem{jp} J. Polchinski, ``Dirichlet-branes and Ramond-Ramond charges,'' 
Phys. Rev. Lett. {\bf 75} (1995) 4724, [hep-th/9510017].
\bibitem{rm} R. C. Myers, ``Dielectric-branes,'' JHEP {\bf 12} (1999) 022, 
[hep-th/9910053].
\bibitem{ds} D. Sorokin, ``Coincident (super)-Dp-branes of codimension 
one, '' JHEP {\bf 08} (2001) 022, [hep-th/0106212]; ``Space-time symmetries and supersymmetry of coincident D-branes,'' Fortsch. Phys. {\bf 50}, (2002), 973. 
\bibitem{nro} Neil R. Constable, Robert C. Myers and Oyvind Tafjord, 
``Fuzzy Funnels: Non-abelian Brane Intersections,'' [hep-th/0105035]; 
``Non-abelian Brane Intersections,'' [hep-th/0102080]; ``The Noncommutative 
Bion Core,'' [hep-th/9911136].
\bibitem{dtv} S. Das, S. Trivedi and S. Vaidya, ``Magnetic Moments of 
Branes and Giant Gravitons,''[hep-th/0008203]; `` Fuzzy Cosets and their 
Gravity Duals,'' [hep-th/0007011].
\bibitem{bds} E.A. Bergshoeff, M. de Roo and A. Servin, ``Towards a 
supersymmetric non-abelian Born-Infeld theory,'' Int. J. Mod. Phys. 
{\bf A 16} (2001) 750, [hep-th/0010151]; ``Non-abelian Born-Infeld 
kappa-symmetry,'' hep-th/0011018; ``On the supersymmetric non-abelian 
Born-Infeld action,'' Fortsch. Phys. {\bf 49} (2001) 433, 
[hep-th/0011264]; E.A. Bergshoeff, A. Bilal, M. de Roo and A. Servin, 
``Supersymmetric non-abelian Born-Infeld revisited,'' JHEP {\bf 07} 
(2001) 029, [hep-th/0105274].
\bibitem{wv} W. Taylor IV, M. Van Raamsdonk, ``Multiple D0-branes in 
weakly curved backgrounds,'' Nucl. Phys, {\bf B 558} (1999) 63, 
[hep-th/9904095]; ``Multiple Dp-branes in weak background fields, '' Nucl. 
Phys, {\bf B 573} (2000) 703, [hep-th/9910052].
\end{enumerate}
\end{document}